
\overfullrule=0Pt.
\input phyzzx.tex
\tolerance=500
\pubtype={ }
\titlepage
\title{A Theorem on Light-Front Quantum Models
\footnote*{This work supported by the U.S. Department of Energy.}}
\author{Wayne N. Polyzou}
\address{Department of Physics and Astronomy \break
The University of Iowa \break
Iowa City, Iowa \quad 52242}
\abstract
I give a sufficient condition for a relativistic front-form quantum
mechanical model to be scattering equivalent to a relativistic
front-form quantum model with an interaction-independent front-form spin.
\submit{J. Math. Phys.}
\endpage
\tenpoint
\chapter{\bf Introduction}

In this paper I present a sufficient condition for a relativistic
front-from quantum model with an interaction dependent spin operator to
be scattering equivalent to a relativistic front-from quantum model with
an interaction independent spin.

Two quantum mechanical models are scattering equivalent if they are
unitarily equivalent and have the same scattering matrix.  Thus,
scattering equivalent models represent equivalent
representations of a given physical system.   In formulating models it is
desirable to work in a representation that has simplifying features.  An
undesirable feature of relativistic light-front quantum mechanics is
that the front-form Hamiltonian and two components of the total angular
momentum necessarily involve interactions.  This complication is
unavoidable.  However, if the angular momentum generators are replaced by
the spin operator it does not follow that the spin operator is
interaction dependent.  In some applications the spin operator
naturally involves interactions, but it is also possible to construct models
with an interaction independent spin operator.

The advantage of working in a representation with a non-interacting spin
operator is that it is straightforward to make approximations that
preserve the relativistic invariance.  For models with an interacting
spin the problem of making approximations that preserve the relativistic
invariance is a non-linear problem.  This non-linearity is the problem
that makes it difficult to find relativistically invariant truncations
of front-form field theories.

The motivation for seeking scattering equivalences with models having an
non-interacting spin operator is to eliminate the difficulty in making
relativistically invariant approximations to more general front-form
models.  In addition, the use of phenomenological models with a
non-interacting front-form spin is justified when the existence of such a
scattering equivalence can be established.

In this paper I prove a theorem that gives sufficient conditions for the
existence of a scattering equivalence that relates front-form models with an
interaction dependent spin to front-form models with an interaction independent
spin.  The theorem proved in this paper applies to systems where particle
number is conserved.  The proof uses standard methods based on two-Hilbert
space scattering theory \Ref\ReedSim{M. Reed and B. Simon, {\it Methods of
Mathematical III, Scattering Theory} (Academic Press, New York, 1979),
p.34-37.}
with simple estimates in the strong topology.  Thus, to the extent that the
abstract structure is preserved, the proof can be extended to settings where
particle number is not conserved.  In models with a production vertex there are
other fundamental issues that need to be addressed such as the definition of an
elementary particle and the definition of the kinematic representation of the
Poincar\'e group.  These issues are not considered in this paper.

A brief description of front-form quantum mechanics is given below.
Section two introduces the notation used in the formulation of the
multichannel scattering theory.  The theorem is stated and proved in
section 3.

A relativistic quantum mechanical model is a model of a system of
particles formulated on a model Hilbert space with the dynamics given by
a unitary representation, $U(\Lambda ,x)$, of the Poincar\'e group.  In the
representation $U(\Lambda ,x)$
$\Lambda$ denotes a Lorentz transformation and $x$ denotes the displacement of
a space-time translation.
This representation defines the dynamics of the model since the
Poincar\'e group contains the time-evolution subgroup.  In the absence
of interactions the dynamics of this system is given by another
representation of the Poincar\'e group, $U_0 (\Lambda ,x)$, called the
kinematic (non-interacting) representation.  Subgroups on which the
interacting and kinematic representations are identical are called
kinematic subgroups.  Dirac \Ref\Dirac{P. A. M. Dirac, Rev. Mod.  Phys.
{\bf 21},392(1949).} investigated the three largest kinematic subgroups.
The largest kinematic subgroup is the 7 parameter subgroup of the
Poincar\'e group that maps the light-front, $x^+ = x^0+x^3 =0$, into
itself.  Relativistic models with this kinematic subgroup are called
front-form models.

Front-form models have three independent one-parameter subgroups that
necessarily involve interactions.  They can be taken as the
one-parameter subgroups that generate translations normal to the light
front and rotations about two independent space axes tangent to the
light front.   Although the total angular momentum operator is
necessarily interaction dependent in front-form models,  it is possible
to satisfy the commutation relations of the Poincar\'e Lie algebra with
a kinematic total front-form
spin operator.\Ref\Fcwp{F. Coester and W.  Polyzou,
Phys.  Rev. D {\bf 26},1348(1982).}\Ref\Bkwp{B. Keister and W.  Polyzou,
{\it Advances in Nuclear Physics, Vol. 20}, edited by J. Negele and E.  Vogt,
(Plenum Publishing, New York, 1991.}  Phenomenological relativistic models with
a kinematic
front-form spin have been employed to model many systems \Ref\Berger{E.
L.  Berger and F. Coester, Phys. Rev. C {\bf 33}, 709(1986).}
\Ref\Glockle{W.  Gl\"ockle, T.-S. H. Lee, and F.  Coester, Phys. Rev. C
{\bf 33}, 709(1986).}\Ref\Polyzoua{W. N.  Polyzou, J. Comp. Phys. {\bf
70}, 117(1987).} \Ref\Chunga{P. L.  Chung, F. Coester, B. D.  Keister,
W. N. Polyzou, Phys. Rev. C {\bf 37}, 2000(1988).} \Ref\Keister{B.  D.
Keister, Phys. Rev. C{\bf 37}, 1765(1988).} \Ref\Chungb{P.L.Chung, F.
Coester, and W.  Polyzou,  Phys.  Lett.  {\bf B}205,545
(1988).}\Ref\Chungc{P.L.Chung, B.D.Keister, F.  Coester, Phys. Rev.
C{\bf 39}, 1544(1989).} where Poincar\'e invariance is an important
symmetry.

There are two relevant examples of representations of the Poincar\'e group with
an interaction dependent spin operator.  The first example is light-front
quantum field theory.   The second example $^{^{[\Fcwp ][\Bkwp]}}$ is any
representation generated by taking tensor products of interacting
representations that have kinematic spin operators.   Tensor products of
interacting representations with kinematic spins have an interacting spin and
satisfy the conditions of the theorem.  The field theory case is interesting
because, due to the infinite number of degree of freedom,  the interacting and
kinematic representation of the Poincar\'e group may act on different Hilbert
spaces\Ref\Haaf{R. Haag, Kgl. Danske Videnskab Selskab Mat. Fys. Medd. {\bf
29},
1(1955).}.  Thus any application to the field theoretic case must be coupled
with a reduction to a finite number of degrees of freedom. Because of this, a
complete treatment of the field theoretic case is beyond the scope of this
paper.

\chapter{\bf Multi Channel Scattering Theory}

The Hilbert space ${\cal H}$ for a system of $N$ particles is the $N$-fold
tensor product of one-particle Hilbert spaces.  The transformation
properties of this system are defined by a representation, $U(\Lambda
,x)$, of the Poincar\'e group on ${\cal H}$.  The dynamics is given by
the time-evolution subgroup, which I denote by
$$T(t) = U[I,
(t,0,0,0)].
\eqn\AA $$
If the $N$-particles are partitioned into a $k$-cluster partition $a$
and the interactions between the particles in different clusters are set
to 0, then the representation $U(\Lambda ,x)$ becomes a new representation
$U_a (\Lambda ,x)$.  This representation physically corresponds to a set
of subsystems that do not interact with each other and, for a $k$-cluster
partition $a$, it can be
represented as the diagonal of the tensor product of $k<N$ subsystem
representations:
$$U_a (\Lambda ,x) := U_{a_1}(\Lambda ,x) \otimes \cdots \otimes
U_{a_k}(\Lambda ,x).
\eqn\AB $$
The $N$-cluster partition corresponds
to turning off all of the interactions.  The representation of the
Poincar\'e group associated with the $N$-cluster partition is the
diagonal of the tensor product of $N$ one-particle representations:
$$U_0 (\Lambda ,x) := \underbrace{U_1(\Lambda ,x) \otimes \cdots \otimes
U_1(\Lambda ,x)}_{\hbox{N-times}}.
\eqn\AC $$

The representations of $U(\Lambda ,x)$ and $U_0(\Lambda ,x)$ are
identical when $(\Lambda ,x)$ is an element of a kinematic subgroup.
It follows that for each partition $a$,
$$U_a (\Lambda ,x)=
U_0(\Lambda ,x)
\eqn\ACAA $$
for $(\Lambda ,x)$ in the kinematic subgroup.

The infinitesimal generators of the Poincar\'e group are the four momentum
$P_{\mu}$ and the antisymmetric angular momentum tensor
$M_{\mu\nu}$.  The spin is related to the Pauli-Lubanski vector which is
defined by
$$W^{\mu} = {1 \over 2}\epsilon^{\mu \alpha \beta \gamma}P_{\alpha}
M_{\beta \gamma}.
\eqn\AD $$
The Poincar\'e group has two independent invariant polynomials in the
infinitesimal generators,
$$M^2 = - P^{\mu}P_{\mu} \qquad W^2 = W^{\mu}W_{\mu} ,
\eqn\AE $$
where  the spectrum of $M^2$ has a positive lower bound and the spin
$j^2$ is defined by
$$j^2 = W^2 / M^2 .
\eqn\AF $$
There are many spin vectors, $\vec{j}$, which are functions of
the generators satisfying $SU(2)$ commutation relations and the relation
$\vec{j}\cdot \vec{j} = W^2 /M^2$.  In front-form models
the spin vector is taken as the front-form spin $^{[\Fcwp]}$.

A single notation is used to treat bound states and scattering states.
Bound states are treated as one fragment scattering channels.

$N$-particle bound states are simultaneous eigenstates of $M^2$ and $W^2$
with eigenvalues in the point spectrum of both of these operators.
These eigenstates can be expressed as linear superpositions of simultaneous
eigenstates of the mass, the light-front components $({\bf p}:= p^+
,p^1,p^2)$ of the four momentum, the spin, and the z-component of
the front-form spin:
$$\vert f_\alpha \rangle = \int dp^+ d^2 p_{\perp} \sum_{\mu=-s}^s \vert
m,s,d,{\bf
p},\mu \rangle  f({\bf p} ,\mu)
\eqn\AG $$
where $f ({\bf p} ,\mu)$ is square
integrable, and $d$ is a degeneracy quantum number.  The degeneracy
quantum number represents relativistically
invariant internal quantum numbers, such as isospin,
that distinguish different types of bound states with the same mass and
spin.  The notation $\alpha$ is used to denote the collection of quantum
numbers $\lbrace m,s,d \rbrace$ and each distinct $\alpha$ is called a
one-body channel.  There is a one-body channel associated with each type
of $N$-body bound state.  The square integrable functions $f ({\bf p}
,\mu)$ span a channel Hilbert space ${\cal H}_{\alpha}$.

The generalized eigenstates $ \vert m,s,d,{\bf p}, \mu
\rangle$ define an isometric mapping from the channel space
${\cal H}_\alpha$ to ${\cal
H}$ by:
$$\Phi_\alpha \vert f \rangle := \vert f_\alpha \rangle
\eqn\AH $$
where the normalization
$$\langle m,s,d,{\bf p}', \mu' \vert
m,s,d,{\bf p}', \mu' \rangle = \delta ({\bf p}-{\bf p}'\,)\delta_{\mu
\mu'}
\eqn\AI $$
is assumed.  Because $\Phi_\alpha$ is an isometry with range all of
${\cal H}_{\alpha}$ it follows that
$$U_{\alpha}(\Lambda ,x) := \Phi_{\alpha}^{\dagger} U(\Lambda ,x)
\Phi_{\alpha}
\eqn\AI $$
defines a unitary representation of the Poincar\'e group
on ${\cal H}_{\alpha}$.

Channels with more than one fragment are associated with scattering
states.  Scattering states $\vert \Psi_\alpha^{\pm}  (t) \rangle $ are
solutions of the time-dependent Schr\"odinger equation that satisfy
either the outgoing $(+)$ or incoming $(-)$ wave asymptotic condition
$$\lim_{t \to \pm \infty} \Vert \, \vert \Psi_\alpha^{\pm}  (t) \rangle - \vert
\phi_\alpha
(t) \rangle \Vert =0 ,
\eqn\AJ $$
where $\vert \phi_\alpha (t) \rangle$ represents a system of
non-interacting particles and/or bound fragments and $\alpha$
distinguishes different scattering channels.  In order to precisely
define a channel, note that the distinct bound fragments define a
partition $a$ of the $N$ particles into $n_a$ clusters, where two
particles are in the same cluster of partition $a$ if they are in the
same asymptotic fragment.  For the partition $a$ to be associated with a
scattering channel each fragment should either have only one particle or
be in a bound state of the subsystem of particles in the fragment.  The
scattering state can be labelled by the quantum numbers of each bound particle
or fragment, $\lbrace m_1,s_1,d_1,{\bf p}_1, \mu_1, \cdots
m_{n_a},s_{n_a},d_{n_a},{\bf p}_{n_a}, \mu_{n_a} \rbrace$. The subset of
quantum numbers $\lbrace m_1,s_1,d_1, m_2,s_2,d_2, \cdots
m_{n_a},s_{n_a},d_{n_a} \rbrace$, which do not include the momenta and magnetic
quantum numbers of the particles,  label the $n_a$ fragment scattering channels
which are denoted by $\alpha$.

With this definition of channel, if some of the particles are identical, there
is a distinct channel for each permutation of particles that changes the
partition $a$ to a partition $a'\not= a$. Channels that differ by the exchange
of identical particles are not physically distinguishable.  In this paper these
channels are treated as being distinguishable with the understanding the cross
section are computed by averaging over equivalent initial channels and summing
over equivalent final channels. The individual bound states and bound clusters
are assumed to have the required symmetry under the exchange of identical
particles.

With this definition, particles are formally treated as distinguishable.
To each channel $\alpha$ there is a unique partition $a=a(\alpha )$
of the system into asymptotic fragments.  It can happen that different
channels $\alpha$ and $\beta$ are associated with same partition; i.e.
$a(\alpha) = a(\beta
)$.  It can also happen that there are no channels
associated with some partitions,  such
as partitions containing two-neutron or two-proton
clusters.

Given a scattering channel, the generalized eigenstates of the four
momentum and 3-component of the front-form spin for each bound fragment
define a mapping from the tensor product of the $k$-bound state channel
spaces
$${\cal H}_{\alpha} := {\cal H}_{\alpha_1} \otimes \cdots \otimes {\cal
H}_{\alpha_2}
\eqn\AK $$
to ${\cal H}$ where each ${\cal H}_{\alpha_i}$ is the space of
square integrable
functions of the front-form momenta and magnetic quantum number of the $i^{th}$
bound fragment.   The mapping, $\Phi_{\alpha}$, is defined by
$$\Phi_{\alpha}\vert f_1 \cdots f_k \rangle := \sum_{\mu_1 \cdots
\mu_k}\int d{\bf p}_1 \cdots d{\bf p}_k \vert
m_1,s_1,d_1,{\bf p}_1, \mu_1 \rangle \otimes \cdots \otimes \vert
 m_k,s_k, d_k,{\bf p}_k, \mu_k \rangle \times $$
$$f_1({\bf p}_1
,\mu_1)\cdots f_k({\bf p}_k ,\mu_k) .
\eqn\AL $$
There is a natural representation of the Poincar\'e group on the channel
space ${\cal H}_{\alpha}$ given by:
$$U_{\alpha}(\Lambda ,x):= U_{\alpha_1}(\Lambda ,x)\otimes \cdots
\otimes U_{\alpha_k}(\Lambda ,x)
\eqn\AM $$
where $U_{\alpha_i}(\Lambda ,x)$ is the irreducible representation of
the Poincar\'e group associated with the bound system of particles in
the i$^{\hbox{\fiverm th}}$ cluster of $a$ as defined by \AI .

It is a consequence of \AL , \AG , and \AI\ that
$$U_{a(\alpha )} (\Lambda ,x) \Phi_\alpha = \Phi_{\alpha} U_\alpha
(\Lambda ,x) .
\eqn\AN $$

The scattering asymptote corresponding to the channel $\alpha$ in
\AJ\ is given by
$$\vert \phi_{\alpha} (t) \rangle := T_{a(\alpha )}(t) \Phi_\alpha  \vert f_1
\cdots f_k \rangle =\Phi_\alpha T_{\alpha} (t) \vert f_1
\cdots f_k \rangle .
\eqn\AO $$
where $T_{a(\alpha )}(t)$  and $T_{\alpha} (t)$ are the time evolution
subgroups of
$U_{a(\alpha )} (\Lambda ,x)$ and $U_{\alpha} (\Lambda ,x)$ respectively.
The asymptotic condition can be reformulated for all channels
simultaneously (trivially
including all $N$-particle bound states) in a two Hilbert space
language.  The asymptotic Hilbert space, ${\cal H}_f$,
is the direct sum of all channel Hilbert spaces:
$${\cal H}_f := \bigoplus_{\alpha} {\cal H}_{\alpha}.
\eqn\AP $$
The sum of the channel injection operators, $\Phi_{\alpha}$
defines a mapping from
${\cal H}_f$ to ${\cal H}$ by:
$$\Phi := \sum_{\alpha} \Phi_{\alpha}
\eqn\AQ $$
where each $\Phi_{\alpha}$ in the sum is understood to act on the
subspace ${\cal H}_{\alpha}$ of ${\cal H}_f$.
The asymptotic representation of the Poincar\'e group on
${\cal H}_f$ is defined by
$$U_f (\Lambda ,x) := \sum_{\alpha} U_{\alpha} (\Lambda ,x)
\eqn\AR $$
where each $U_{\alpha}(\Lambda ,x)$ acts on the subspace
${\cal H}_\alpha$.
Scattering solutions, $\vert \Psi^{\pm} (0) \rangle$,
are defined in terms of two-Hilbert space wave operators which
are mappings from ${\cal H}_f$ to ${\cal H}$ defined by:
$$\vert \Psi^{\pm} (0) \rangle :=
\Omega_{\pm} (T, \Phi, T_f) \vert \phi_f \rangle
\eqn\AQAA $$
$$\Omega_{\pm} (T, \Phi, T_f) := s-\lim_{t \to \pm \infty} T(t) \Phi T_f
(-t)
\eqn\AS $$
where $T(t)$ and $T_f(t)$ are the time evolution subgroups of $U(\Lambda
,x)$ and $U_f (\Lambda ,x)$ respectively, and $\vert \phi_f \rangle$ is
a vector in ${\cal H}_f$.  A similar notation is used
for the individual channel scattering
state vectors and channel wave operators
$$\vert \Psi^{\pm}_\alpha (0) \rangle :=
\Omega_{\pm} (T, \Phi, T_f) \vert \phi_\alpha \rangle =
\Omega_{\alpha \pm} \vert \phi_\alpha \rangle
\eqn\AQAB $$
$$\Omega_{\alpha \pm} := \Omega_{ \pm} (T, \Phi_\alpha, T_\alpha)
:= s-\lim_{t \to \pm \infty} T(t) \Phi_\alpha  T_\alpha
(-t)
\eqn\ASAA $$
where $\vert \phi_\alpha \rangle$ is a vector in the channel Hilbert
space ${\cal H}_{\alpha}$.

The scattering operator is the following mapping from ${\cal
H}_f$ to itself:
$$S := \Omega_{+}^{\dagger} (T, \Phi, T_f)\Omega_{-} (T, \Phi, T_f)  .
\eqn\AT $$

The scattering operator will be unitary if
$$\Omega_{+}(T, \Phi, T_f)\Omega_{+}^{\dagger} (T, \Phi, T_f)  =
\Omega_{-}(T, \Phi, T_f)\Omega_{-}^{\dagger} (T, \Phi, T_f),
\eqn\AU $$
which is the property that the wave operators are asymptotically
complete.

The wave operators are Poincar\'e invariant provided they satisfy the
intertwining relations:
$$U(\Lambda ,x) \Omega_{\pm}(T, \Phi, T_f) =  \Omega_{\pm}(T, \Phi, T_f)
U_f (\Lambda ,x)
\eqn\AV $$
for all Poincar\'e transformations.  This implies that the scattering
operator is Poincar\'e invariant
$$[U_f (\Lambda ,x) , S ]_{-}=0.
\eqn\AW $$

If a representation of the Poincar\'e group is transformed by
a unitary transformation, it does not follow that the original theory and
the transformed theory have the same scattering operator.  Unitary
transformations that preserve the scattering operator are called
scattering equivalences.  In the two Hilbert space setting a sufficient
condition for $U'(\Lambda ,x)= A^{\dagger} U(\Lambda ,x)A$ to be
scattering equivalent to $U(\Lambda ,x)$ is $^{^{[\Fcwp]}}$
$$\lim_{t \to \pm \infty} \Vert (A-1 )\Phi T_f (-t) \vert \xi \rangle
\Vert = 0
\eqn\AX $$
for {\it both} time limits.  An operator $A$ satisfying \AX\ is said to
be {\it asymptotically equivalent to the identity with respect to
$\Phi$}.
When \AX\ holds it follows that
$\Omega_{\pm} ( T,A\Phi,T_f ) = \Omega_{\pm} ( T,\Phi,T_f ) $
which implies:
$$\Omega'_{\pm} = \Omega_{\pm} ( T',\Phi,T_f ) = \Omega_{\pm} (
A^{\dagger}TA,\Phi,T_f ) = A^{\dagger} \Omega_{\pm} ( T,A\Phi,T_f ) =
A^{\dagger} \Omega_{\pm} ( T,\Phi,T_f ) = A^{\dagger} \Omega_{\pm}
\eqn\AY $$
and
$$S' = \Omega^{\prime \dagger}_{+} \Omega^{\prime}_- =
\Omega^{ \dagger}_{+}A A^{\dagger} \Omega_- =
\Omega^{\dagger}_{+} \Omega_- = S.
\eqn\AZ $$

\chapter{\bf Statement and Proof of the Theorem}

In this section sufficient conditions for the existence of a scattering
equivalence between a front form-model with an interaction dependent spin and a
front-form model with an interaction independent spin are established by
proving
the following theorem:

\noindent {\bf Theorem:} {\it Let $U(\Lambda ,x)$ be the representation
of the Poincar\'e group for a model of a system of $N$ interacting
particles and let $U_a (\Lambda ,x)$ be the representation obtained from
$U(\Lambda ,x)$ by turning off the interactions between particles in
different clusters of the partition $a$.  Assume that $U(\Lambda ,x)$
(and consequently $U_a (\Lambda ,x)$) has the kinematic subgroup of the
light-front and that the model satisfies:

\item{1.} The wave operators exist and are asymptotically complete .

\item{2.} The wave operators are Poincar\'e invariant .

\item{3.} There exits unitary operators $A_a$ on ${\cal H}$ that are
kinematically invariant and satisfy

\itemitem{a.)} $\lim_{t \to \pm \infty} \Vert
(A_{a(\alpha )} -1)\Phi_{\alpha}T_\alpha (t) \vert \xi_\alpha \rangle
\Vert =0$
\itemitem{}{\it and }
\itemitem{b.)}The operator $\sum_s P(s) A_a P_a(s)$ has a bounded inverse for
each partition $a$ of $N$-particles, where $P(s)$ and $P_a(s)$ are the
orthogonal projectors on the invariant spin $s$ subspaces associated
with the representations $U(\Lambda ,x)$ and $U_a (\Lambda ,x)$
respectively.

\noindent Under these conditions the representation $U(\Lambda ,x)$ is
scattering equivalent to a representation $\bar{U}(\Lambda ,x)$ with the
kinematic subgroup of the light-front and with a kinematic front-form
spin operator.}

The first two assumptions of the theorem are sufficient to ensure the
model has a reasonable physical interpretation.  The first assumption
implies the unitarity of the scattering matrix while the second implies
the Poincar\'e invariance of the scattering matrix.

The third assumption is the nontrivial assumption.  It is a mild
condition because of the freedom available to choosing the operators
$A_a$.  The simplest choice is to choose $A_a=1$.  This choice is
appropriate for models that already have a non-interacting $j^2$.  It is
also appropriate for models with interacting spins provided that
$\sum_s P(s) P_a(s)$ remains invertible.  If the sum is not
invertible, then the theorem permits modifications of the condition by the
insertion of an operator $A_a$ between the projectors.  For the case of tensor
products of models with non-interacting spins, it is know that if $A_a$ is
taken as $A_a = B^{\dagger} B_a$ where $B$ and $B_a$ are the Sokolov or packing
operators for the models $^{^{[\Fcwp][\Bkwp]}}$, then the third condition of
the theorem is satisfied,  although in this example it is not known whether
condition 3b.) is satisfied for  $A_a=1$.

The proof of the theorem is based on a number of lemmas.

By assumption the wave operators,
$\Omega_{\pm}(T,\Phi,T_f)$, exist as unitary mappings from
${\cal H}_f$ to ${\cal H}$, are
asymptotically complete,
$$\Omega_{+}(T, \Phi, T_f)\Omega_{+}^{\dagger} (T, \Phi, T_f)  =
\Omega_{-}(T, \Phi, T_f)\Omega_{-}^{\dagger} (T, \Phi, T_f),
\eqn\BA $$
and Poincar\'e invariant,
$$U(\Lambda ,x) \Omega_{\pm}(T, \Phi, T_f) =  \Omega_{\pm}(T, \Phi, T_f)
U_f (\Lambda ,x).
\eqn\BB $$
In addition $U(\Lambda ,x)$ has the kinematic subgroup of the light
front, which means
that $U(\Lambda ,x)= U_0 (\Lambda ,x)$ for Poincar\'e transformations
$(\Lambda ,x)$
that leave the null plane $x^+=0$ invariant.

The technical assumption is that the sum
$$X_a := \sum_s P(s) A_a P_a(s)
\eqn\BC $$
has a bounded inverse for each partition $a$ where $P_a(s)$ is the
projection on the invariant subspace of $U_a (\Lambda ,x)$ on which
$j_a^2:= W^2_a /M_a^2$ has eigenvalue $s(s+1)$, and $P(s)$ is the
corresponding projector associated with $j^2$.  The unitary operator
$A_a$ is any kinematically invariant
operator that is asymptotically equivalent to the identity
with respect to $\Phi_{\alpha}$ for all $\alpha$ with $a=a(\alpha )$.
The freedom to choose $A_a$ can be used to make the $X_a$ have a bounded
inverse.

The proof of the theorem follows as a consequence of the lemmas that
follow.

\noindent {\bf Lemma 1:} {\it With $P(s)$, $P_a(s)$, and $A_a$
as defined in the
statement of the theorem
}
$$X_a:= \sum_{s} P(s)A_a P_a(s)
\eqn\BD $$
{\it satisfies}
$$\Vert X_a \Vert \leq 1.
\eqn\BE $$

\noindent {\bf Proof:} Let $\vert \xi (s) \rangle $ denote a unit
normalized vector in the range of $P_a (s)$.  Then
$$\Vert X_a \vert \xi (s) \rangle \Vert  = \Vert \sum_r P(r) A_a P_a(r) \vert
\xi (s) \rangle \Vert = \Vert P(s) A_a \vert
\xi (s) \rangle \Vert \leq \Vert P(s)A_a \Vert \, \Vert \vert \xi (s) \rangle
\Vert = 1 .
\eqn\BF $$
Next observe that any normalizable state can be expanded in the form
$$\vert \xi \rangle = \sum_s c_s \vert \xi (s) \rangle  \qquad
\hbox{with} \qquad \sum_s \vert c_s \vert^2 < \infty
\eqn\BG $$
where $\vert \xi (s) \rangle$ are unit normalized vectors in the range of
$P_a(s)$.  The orthogonality of the $P(s)$'s imply
$$\Vert X_a \vert \xi \rangle \Vert = \sum_s \vert c_s \vert^2 \, \Vert
P(s)A_a
\vert \xi (s) \rangle \Vert^2 \leq \sum_s \vert c_s \vert^2 = \Vert
\vert \xi
\rangle \Vert .
\eqn\BH $$
The lemma follows by dividing the left side of the equation by the
right.

The next lemma is the key result that is needed to establish scattering
equivalence with a system with non-interacting $j^2$.

\noindent {\bf Lemma 2:} {\it For
$\vert \xi\rangle \in {\cal H}_{\alpha}$
the assumptions of the theorem imply}
$$\lim_{t \to \pm \infty} \Vert [X_a-1] \Phi_\alpha T_\alpha (-t)
\vert \xi \rangle \Vert =0
\eqn\BI $$
{\it and}
$$\lim_{t \to \pm \infty} \Vert [X_a^{\dagger}-1] \Phi_\alpha T_\alpha (-t)
\vert \xi \rangle
\Vert =0.
\eqn\BJ $$

\noindent {\bf Proof:} By the Poincar\'e invariance of the wave
operators the wave operators intertwine the asymptotic and interacting spin
operators:
$$j^2 \Omega_{\alpha \pm} = \Omega_{\alpha \pm} j_\alpha^2 .
\eqn\BK $$
This implies
$$P(s)  \Omega_{\alpha \pm} = \Omega_{\alpha \pm} P_\alpha(s)
\eqn\BL $$
for each value of $s$, where $P_\alpha (s)$ it the projector on the
invariant spin $s$ subspace associated with the representation $U_\alpha
(\Lambda ,x)$ of ${\cal H}_{\alpha}$.
For $Q(s):= [1  - P(s)]$ the orthogonality of $P(s)$ and $Q(s)$ implies
$$0 = Q(s) P(s) \Omega_{\alpha \pm} = Q(s) \Omega_{\alpha \pm} P_\alpha (s).
\eqn\BM $$
Using the definition of the channel wave operator \ASAA\ in \BM\ and
letting it act on a spin $s$ channel state,
$\vert \xi \rangle = P_{\alpha}(s) \vert \xi \rangle$, gives
$$0 = \lim_{t \to \pm \infty} \Vert Q(s) T(t) \Phi_\alpha
T_\alpha (-t) \vert \xi \rangle \Vert = $$
$$\lim_{t \to \pm \infty} \Vert Q(s) T(t) A_{a(\alpha )} \Phi_\alpha
T_\alpha (-t) \vert \xi \rangle \Vert = $$
$$\lim_{t \to \pm \infty} \Vert [1 - P(s)] T(t) A_{a(\alpha )}
P_{a(\alpha )}(s)
\Phi_\alpha T_\alpha (-t) \vert \xi \rangle \Vert = $$
$$\lim_{t \to \pm \infty} \Vert [1 - P(s)] A_{a(\alpha )} P_{a(\alpha )}
(s) \Phi_\alpha
T_\alpha (-t) \vert \xi \rangle \Vert =
\eqn\BN $$
$$\lim_{t \to \pm \infty} \Vert [1 - \sum_{r}P(r)A_{a(\alpha )}
P_{a(\alpha )}(r)]
\Phi_\alpha T_\alpha (-t) \vert \xi \rangle
\Vert = $$
$$\lim_{t \to \pm \infty} \Vert [1 - X_{a(\alpha )}]
\Phi_\alpha T_\alpha (-t) \vert \xi \rangle \Vert
\eqn\BO $$
where the relation $[P(s),T(t)]_-=0$, equation \AN\ and
the asymptotic equivalence
$$\lim_{t \to \pm \infty} \Vert (A_{a(\alpha )} - I)  \Phi_\alpha
T_\alpha (-t) \vert \xi \rangle \Vert = 0
\eqn\BOAA $$
were used to obtain \BO .  This proves the first part of lemma 2.  To
prove the second part note
$$\lim_{t \to \pm \infty} \Vert [1 - X_{a (\alpha )}^{\dagger}]
\Phi_\alpha T_\alpha (-t) \vert \xi \rangle \Vert = $$
$$\lim_{t \to \pm \infty} \Vert [1 - \sum_{r}P_{a(\alpha )
}(r)A^{\dagger}_{a(\alpha )}
P(r)]
\Phi_\alpha T_\alpha (-t) \vert \xi \rangle \Vert = $$
$$\lim_{t \to \pm \infty} \Vert [P_{a(\alpha )}(s)
A^{\dagger}_{a(\alpha )}A_{a(\alpha )}
 - \sum_{r}P_{a(\alpha )} (r)
A^{\dagger}_{a(\alpha )}P(r)P_{a(\alpha )}(s)]
\Phi_\alpha T_\alpha (-t) \vert \xi \rangle \Vert \leq $$
$$\lim_{t \to \pm \infty} \bigl [
\Vert \sum_{r \not= s} P_{a(\alpha )}(r)
A^{\dagger}_{a(\alpha )}P(r)Q(s) P_{a(\alpha )}(s)
\Phi_\alpha T_\alpha (-t) \vert \xi \rangle \Vert + $$
$$
\Vert P_{a(\alpha )}(s) A^{\dagger}_{a(\alpha )} (P(s)-
A_{a(\alpha )}) P_{a(\alpha )}(s)
\Phi_\alpha T_\alpha (-t) \vert \xi \rangle \Vert \bigr ] \leq $$
$$\lim_{t \to \pm \infty} \bigl [
\Vert \sum_{r \not= s} P_{a(\alpha )}(r)
A^{\dagger}_{a(\alpha )}
P(r)Q(s) \Vert +
\Vert P_{a(\alpha )}(s)
A^{\dagger}_{a(\alpha )}
\Vert \bigr ]  \Vert (P(s)- 1) P_{a(\alpha )}(s)
\Phi_\alpha T_\alpha (-t) \vert \xi \rangle \Vert
\eqn\BP $$
which vanishes by \BN\ ( the coefficient is bounded by lemma 1).
Equation \BOAA\ was used again in \BP .  The completes the proof of lemma 2.

The operators $X_{a(\alpha )}\Phi_{\alpha}$ and $X_{a (\alpha )}
^{\dagger} \Phi_{\alpha}$ are
suitable injection operators for a scattering theory but neither one is
unitary.
The next lemma provides a unitary injection operator
constructed out of $X_a$ and $X_a^{\dagger}$.

\noindent {\bf Lemma 3:} {\it For $\vert \xi \rangle \in {\cal H}_{\alpha}$:
the assumptions of the theorem imply:}
$$\lim_{t \to \pm \infty} \Vert [Y_{a(\alpha )} -1]
\Phi_\alpha T_\alpha (-t) \vert \xi \rangle
\Vert =0
\eqn\BQ $$
{\it where}
$$Y_{a( \alpha )}:= (X_{a (\alpha )}X_{a (\alpha )}^{\dagger})^{-1/2}X_{a
(\alpha )}.
\eqn\BR $$

\noindent {\bf Proof:} For the proof of this theorem
the notation $a$ is used as an abbreviation for $a(\alpha )$.
Since $X_a$ is bounded by lemma 1 and has bounded
inverse by assumption it follows that
$(X_a^{\dagger})^{-1}=(X_a^{-1})^{\dagger}$ exists and that
$X_aX_a^{\dagger}$ is a bounded positive operator with bounded inverse.
Consequently, $(X_aX_a^{\dagger})^{-1/2}$ exists and is bounded with
bounded inverse.  To prove the theorem I consider the inequalities:

$$\Vert [Y_a -1] \Phi_\alpha T_\alpha (-t) \vert \xi \rangle \Vert =$$
$$\Vert [(X_aX_a^{\dagger})^{-1/2}X_a  -1] \Phi_\alpha T_\alpha
(-t) \vert \xi \rangle \Vert
\leq $$
$$\Vert [(X_aX_a^{\dagger})^{-1/2}\Vert \, \Vert [X_a  -1] \Phi_\alpha
T_\alpha (-t) \vert \xi \rangle \Vert + \Vert [(X_aX_a^{\dagger})^{-1/2}
-1] \Phi_\alpha T_\alpha (-t) \vert \xi \rangle \Vert \leq $$
$$\Vert [(X_aX_a^{\dagger})^{-1/2}\Vert \, \Vert [X_a  -1]
 \Phi_\alpha T_\alpha (-t) \vert \xi \rangle \Vert +
\Vert [(X_aX_a^{\dagger})^{-1/2} {1 \over 1 +
(X_aX_a^{\dagger})^{1/2}} \Vert \, \Vert [X_aX_a^{\dagger} -1]
\Phi_\alpha T_\alpha (-t) \vert \xi \rangle \Vert  .
\eqn\BS $$
As $t\to \pm \infty$ the first term vanishes
by lemma 2 and the boundedness of
$(X_aX_a^{\dagger})^{-1/ 2}$.  The second term is bounded by:
$$\Vert [(X_aX_a^{\dagger})^{-1/2} {1 \over 1 +
(X_aX_a^{\dagger})^{1/2}} \Vert \, \Vert [X_aX_a^{\dagger} -1]
\Phi_\alpha T_\alpha (-t) \vert \xi \rangle \Vert \leq $$
$$\Vert [(X_aX_a^{\dagger})^{-1/2} {1 \over 1 +
(X_aX_a^{\dagger})^{1/2}} \Vert \, \bigl [ \Vert X_a \Vert \,
\Vert [X_a^{\dagger} -1] \Phi_\alpha T_\alpha (-t) \vert \xi \rangle
\Vert + \Vert [X_a -1] \Phi_\alpha T_\alpha (-t) \vert \xi \rangle
\Vert \bigr ]
\eqn\BT $$
which also vanishes as $t\to \pm \infty$ by lemma 2.  Note that the
operator norms in the above expression are all finite. This completes
the proof of lemma 3.

The next lemma establishes that $Y_a$ is unitary, commutes with the
kinematic subgroup of the light front, and intertwines $j^2$ and
$j_a^2$:

\noindent{\bf Lemma 4:}{\it Under the assumptions of the theorem the
operator $Y_a$ defined in the previous lemma is unitary, commutes with the
kinematic subgroup of the light front, and satisfies $j^2 Y_a = Y_aj_a^2$.}

\noindent {\bf Proof:}  Unitarity follows by computation:
$$Y_a^{-1}=[(X_aX_a^{\dagger})^{-1/2}X_a]^{-1}
=X_a^{-1}(X_aX_a^{\dagger})^{1/2}=X_a^{\dagger}
(X_a^{\dagger})^{-1}X_a^{-1}(X_aX_a^{\dagger})^{1/2}
=$$
$$X_a^{\dagger}
(X_aX_a^{\dagger})^{-1}(X_aX_a^{\dagger})^{1/2}=
X_A^{\dagger}
(X_AX_a^{\dagger})^{-1/2} = Y_a^{\dagger}.
\eqn\BR $$

The intertwining property also follows by computation:
$$j^2 X_a = j^2 \sum_s P(s) A_{a} P_a(s) =
\sum_s s(s+1) P(s) A_{a} P_a(s) = \sum_s P(s)
A_{a} P_a(s)s(s+1) =$$
$$\sum_s P(s) A_{a(\alpha )}  P_a(s) j_a^2 = X_a j_a^2.
\eqn\BS $$
Similarly it can be shown that $j_a^2 X_a^{\dagger} = X_a^{\dagger} j^2$.
It follows that
$$[ (X_aX_a^{\dagger}),j^2]_-=0 \rightarrow [
(X_aX_a^{\dagger})^{-1/2},j^2]_-=0 .
\eqn\BT $$
Combining \BT\ with \BS\ gives
$$j^2 Y_a = j^2 (X_aX_a^{\dagger})^{-1/2}X_a
=(X_aX_a^{\dagger})^{-1/2}j^2 X_a=
(X_aX_a^{\dagger})^{-1/2} X_a j_a^2 =Y_a j_a^2.
\eqn\BU $$

That $Y_a$ commutes with the kinematic subgroup follows because each
$P(s)$, $P_a (s)$, and $A_a$ commute with the kinematic subgroup.  This
completes the proof of lemma 4.

\noindent {\bf Proof of the Theorem:} The first step in the proof of the
theorem is to construct a scattering equivalence that eliminates the
interaction dependence in $j^2$.  The second step is to construct another
scattering equivalence that removes the interaction dependence in $\hat{z}
\times \vec{j}$.  I follow closely the discussion in $^{[\Fcwp]}$.

To prove the theorem I define the transformed
representation of the Poincar\'e group:
$$U_y ( \Lambda ,x):= Y_0^{\dagger} U(\Lambda ,x ) Y_0
\eqn\BV $$
and the transformed channel injection operators
$$\Phi_{y \alpha} = Y_0^{\dagger} Y_{a (\alpha )} \Phi_{\alpha}
\eqn\BW $$
where $Y_0$ is the $Y_{a(\alpha )}$ corresponding to the $N$-cluster
partition $a=0$.
The full-two Hilbert space injection operator for this representation is
defined by
$$\Phi_y := \sum_{\alpha} \Phi_{y \alpha}
\eqn\BX $$
where the sum runs over all channels including bound (one-cluster
channels).

Lemma three implies that for
$$\bar{\Phi} := \sum_{\alpha} Y_a \Phi_\alpha
\eqn\BXAA $$
that $\Omega_{ \pm}(T  , \Phi , T_f ) = \Omega_{ \pm}(T  , \bar{\Phi} ,
T_f )$ which gives the relation
$$\Omega_{y \pm} := \Omega_{ \pm}(T_y , \Phi_y , T_f ) = \Omega_{
\pm}(Y_0^{\dagger} T Y_0 , Y_0^{\dagger} {\bar \Phi} , T_f ) = Y_0^{\dagger}
\Omega_{ \pm}(T  , \Phi , T_f ) =Y_0^{\dagger} \Omega_{\pm} .
\eqn\BY $$
This equation establishes both the existence of the transformed wave
operators and the scattering equivalence with the original theory.

The transformed injection operators are constructed to satisfy the
intertwining properties similar to \AN :

$$\Phi_y U_f (\Lambda ,x) \vert \xi_{\alpha} \rangle =
\bar{U}_{a(\alpha)} (\Lambda ,x)\Phi_y \vert \xi_{\alpha} \rangle
\eqn\BZ $$
where
$$\bar{U}_a (\Lambda ,x):= Y_0^{\dagger} Y_a U_a (\Lambda ,x) Y_{a}^{\dagger}
Y_0
\eqn\BZAA $$
and where for each partition $a$, $\bar{U}_a (\Lambda ,x)$ is a unitary
representation of the Poincar\'e group with (1) the kinematic subgroup
of the light front and (2) the same spin operator as the noninteracting
system.
The first claim follows because $U_a (\Lambda ,x)$
has the kinematic subgroup of the light front for each $a$ and
$Y_{a}^{\dagger} Y_0$ commutes with the kinematic subgroup by lemma four.
The second claim follows by lemma four because
$$Y_{a}^{\dagger} Y_0 j_0^2 = Y_{a}^{\dagger}j^2  Y_0 =
j_a ^2Y_{a}^{\dagger} Y_0 .
\eqn\CA $$

The asymptotic completeness and Poincar\'e invariance of the wave
operators $\Omega_{y \pm}$ follow from the corresponding properties of
the untransformed wave operators and the relation \BY\ :
$$\Omega_{y\pm}=Y_0^{\dagger} \Omega_{\pm}.
\eqn\CB $$

It follows that $U_y (\Lambda ,x)$ and $U(\Lambda ,x)$ are scattering
equivalent.  This does not complete the proof of the theorem because
although $U_y (\Lambda ,x)$ has the kinematic subgroup of the light-front
and satisfies $j_y^2 = j_0^2$, it does not follow that $\vec{j}_y =
\vec{j}_0$.  Specifically the raising and lowering operators
for the front-form spin may be interaction dependent in this
representation.  Given that this is the case, the Lorentz
invariance of the scattering operator implies

$$[ j_{f \pm} ,\Omega^{\dagger}_{y+} \Omega_{y -} ]_{-}=0
\eqn\CC $$
where $j_{f \pm}$ are the asymptotic raising and lowering operators.
Since the scattering matrix also commutes with $j_f^2$ and $\hat{z}
\cdot \vec{j}_{f}$,
the matrix elements have the form:

$$_f\langle \beta s \mu \vert S \vert \beta' s' \mu' \rangle_f = \delta_{ss'}
\delta_{\mu \mu'}{} _f\langle \beta \vert \hat{S}(s) \vert \beta' \rangle_f
\eqn\CD $$
where $\beta$ denotes the remaining quantum numbers.
The matrix elements of the wave operators between the non-interacting
and asymptotic basis vectors satisfy
$$_0\langle \beta s \mu \vert \Omega_{y\pm} \vert \beta' s' \mu' \rangle_f
= \delta_{ss'}
\delta_{\mu \mu'}{} _0\langle \beta \vert \hat{\Omega}_{\pm}(s,\mu) \vert
\beta' \rangle_f
\eqn\CE $$
because the wave operators intertwine to the $y$ representation, where
$j^2$ and $\hat{z}
\cdot \vec{j}$  are kinematic but
the raising and lowering operators may be non-trivial.  The non-triviality of
the
raising and lowering operators results in the
$\mu$ dependence in matrix elements of the
reduced wave operator $\hat{\Omega}_{\pm}(s,\mu)$.
Equation \CD\ implies that
$$_f\langle \beta \vert \hat{\Omega}_{y+}^{\dagger}  (s,\mu)
\hat{\Omega}_{y-} (s,\mu) \vert \beta' \rangle_f =
_f\langle \beta \vert \hat{\Omega}_{y+}^{\dagger}  (s,s)
\hat{\Omega}_{y-} (s,s) \vert \beta' \rangle_f
\eqn\CF $$
independent of $\mu$.
Asymptotic completeness then implies
$$_0\langle \beta s \mu \vert Z \vert \beta' s' \mu' \rangle_0:=
\delta_{ss'}\delta_{\mu \mu'}{}_0\langle \beta \vert
\hat{\Omega}_{y-}(s,s)\hat{\Omega}_{y-}^{\dagger}  (s,\mu)
\vert \beta' \rangle_0
=\delta_{ss'}\delta_{\mu \mu'}{} _0\langle \beta \vert
\hat{\Omega}_{y+}(s,s)\hat{\Omega}_{y+}^{\dagger}  (s,\mu)
\vert \beta' \rangle_0
\eqn\CG $$
is unitary and independent of the choice of asymptotic condition.
It is useful to introduce the notation
$$_0\langle \beta s \mu \vert \bar{\Omega}_{y \pm} \vert \beta' \mu' s'
\rangle_f :=
\delta_{ss'}\delta_{\mu \mu'}{}_0\langle \beta \vert
\hat{\Omega}_{y\pm}(s,s)\vert
\beta'\rangle_f .
\eqn\CH $$
In this notation \CF\ becomes
$$Z = \bar{\Omega}_{y +} \Omega^{\dagger}_{y +} =
\bar{\Omega}_{y -} \Omega^{\dagger}_{y -}
\eqn\CI $$
from which it follows that
$$\bar{\Omega}_{y \pm} =  Z \Omega_{y \pm} .
\eqn\CJ $$
The structure of $\bar{\Omega}_{y \pm}$ implies that the spin raising
and lowering operators satisfy
$$j_{0 \pm} \bar{\Omega}_{y \pm} = \bar{\Omega}_{y \pm}j_{f \pm}
\eqn\CK $$
since by construction there is no $\mu$ dependence in the kernel of the
matrix elements of
$\bar{\Omega}_{y \pm}$ in the free-asymptotic representation.

The result is that the desired representation is given by
$$\bar{U}_y (\Lambda ,x) := Z Y_0^{\dagger} U(\Lambda ,x) Y_0 Z^{\dagger}
\eqn\CL $$
with channel injection operators given by
$$\bar{\Phi}_{\alpha} = Z Y_0^{\dagger} Y_a \Phi_{ \alpha}
\eqn\CM $$

To show that this system leads to a scattering theory with the desired
properties note the wave operators exist since:
$$\Omega_{y \pm}(\bar{T}_y, \bar{\Phi}_y , T_f )=
\Omega_{y \pm}( Z T_y Z^{\dagger} , Z\Phi_y , T_f )=
Z \Omega_{y \pm}( T_y  , \Phi_y , T_f ) = ZY_0\Omega_{\pm} (T, \Phi ,T_f)
\eqn\CN $$
exists.   The scattering equivalence to the original wave operators is
a consequence of \CN .

These operators are asymptotically complete since
$$\bar{\Omega}_{y +}\bar{\Omega}^{\dagger}_{y +} =
Z \Omega_{y +}\Omega^{\dagger}_{y +} Z^{\dagger}=
Z \Omega_{y -}\Omega^{\dagger}_{y -} Z^{\dagger}=
\bar{\Omega}_{y -}\bar{\Omega}^{\dagger}_{y -}
\eqn\CO $$
and Poincar\'e invariant since
$$\bar{U}(\Lambda ,x)\bar{\Omega}_{y \pm}=
Z U_y (\Lambda ,x) \Omega_{y \pm}=
Z \Omega_{y \pm} U_f (\Lambda ,x) =
\bar{\Omega}_{y \pm} U_f (\Lambda ,x) .
\eqn\CP $$
The kinematic invariance follows because $Z$ and $Y_0$ commute with the
kinematic subgroup.  Finally the intertwining properties of the wave
operators, $Y_0$ and $Z$ ensure that the spin vector is kinematic.  This
completes the proof of the theorem.

\chapter{\bf Conclusion:}

In this paper I have given a sufficient condition for a front-form
quantum model with an interacting spin to be scattering equivalent to a
front-form model with a kinematic front form-spin.  The essential
property is that there exist a set of kinematically invariant unitary
operators $A_a$ such that $A_{a(\alpha )}$ is asymptotically equivalent
to the identity with respect to the channel injection operator
$\Phi_{\alpha}$ and has the property that the operators $X_a= \sum_s P(s)
A_a P_a (s)$, have bounded inverses.

The simplest case is when $\sum_s P(s) P_a(s)$ has a bounded inverse.
In this case all of the operators $A_a$ can be taken to be the identity.
When this fails the freedom to choose the operators $A_a$ can be
utilized, although choice of $A_a$ is limited by the asymptotic conditions.
The theorem is also applicable to the case where the model satisfies
spacelike cluster properties.  In that case if the operators $A_a$ are
taken to be $A_a = B^{\dagger} B_a$ where $B$ and $B_a$ are the Sokolov
or packing operators that map representations that cluster properly to
representation with a kinematic spin, all of the conditions of the
theorem are satisfied.

In general models need to be considered on a case to case basis.  The
ability to choose the operators $A_a$ leaves a fair amount of
flexibility in establishing the scattering equivalence.  As it stands,
the theorem does not directly apply to the interesting case of field
theories.  The basic elements of the proof should extend to the field
theoretic case although it is not known how to control the usual
difficulties that arise in models with an infinite numbers of degrees of
freedom.

The author would lie to thank
\vfill\eject
\refout
\end